\documentstyle[12pt,aasms4]{article}

\begin{document}

\title{Neutrino Production of Deuterium in Supermassive Stars and Possible
Implications for Deuterium Detections in Ly-Limit Systems}

\author{George M. Fuller and Xiangdong Shi}
\affil{Department of Physics, University of California, San Diego, La Jolla,
CA 92093}
\authoremail{gfuller@ucsd.edu, shi@physics.ucsd.edu}

\begin{abstract}
We describe how thermally produced anti-electron neutrinos ($\bar\nu_e$) from
the homologously collapsing core of a supermassive star (${\rm M} \ga
5\times10^4\,{\rm M_\odot}$) can lead to significant deuterium enrichment in
the ejected envelope of such a star.  Deuterium-enriched material at high
redshift might then serve as a clue to the existence of pregalactic
supermassive stars. Conceivably, the ejected deuterium-enriched material could
intercept the line of sight to a distant QSO and mimic a Lyman limit absorber.
In such a case, the deuterium abundance inferred from absorption lines might
not reflect the true primordial abundance of deuterium. 
We discuss relevant theoretical uncertainties in supermassive star physics as
well as potential observational signatures in Lyman $\alpha$ absorber clouds
for processing by stars of this kind.
\end{abstract}

\keywords{elementary particles - nuclear reactions, nucleosynthesis,
abundances - cosmology: observations and theory}

\section{Introduction}

In this paper we assess the prospects for deuterium synthesis resulting from
$\bar\nu_e$-induced neutron production in supermassive stars (${\rm M}\ga
5\times 10^4 {\rm M_\odot}$ - these objects become dynamically unstable due
to the general relativistic instability). We will show
that this process cannot account for the observed interstellar or solar system
deuterium abundance. However, it might result in {\it local} enrichment of
deuterium above its primordial value in the ejecta from a collapsed
supermassive star. This might give a telltale sign of the existence and
evolution history of pre-galactic supermassive stars - as yet there is no
direct evidence for these objects ever having been extant in the universe,
though it has been argued that their formation may be an inevitable consequence
of the evolution of dense star clusters and perhaps ultra-low metallicity
environments in the early universe (Hoyle \& Fowler 1963; Begelman \& Rees
1978; Fuller, Woosley, \& Weaver 1986, hereafter FWW;
Bond, Arnett, \& Carr 1984; McLaughlin \& Fuller 1996).

If such deuterium-enriched ejecta at high redshift intercept a line of sight
to a more distant QSO, then it could mimic a Lyman limit absorber.
Measurement of isotope-shifted hydrogen absorption lines in such a system may
then lead to an erroneous estimate of the primordial deuterium abundance
(D$/$H) and, hence, an incorrect inference of the fraction of the closure
density contributed by baryons, $\Omega_b$. Future indisputable observations of
significantly discordant D$/$H values among different Lyman limit systems, or
very high values of this quantity, may be a signal for the existence of
pregalactic supermassive stars. Confirmation of such a conjecture would depend
on other evidence of supermassive star processing (FWW): nucleosynthesis
products of hot hydrogen burning (the $rp$-process, Wallace \& Woosley 1981);
greatly enhanced helium abundances; or effects of large black holes.

Deuterium is very fragile in stars.  Thus to search for
the primordial value of D$/$H one has to look for gas clouds unprocessed by
stars.  Low metallicity Lyman $\alpha$ absorbers at high redshifts are  
thought to be such primordial clouds with little stellar processing ({\it cf.}
Jedamzik \& Fuller 1997).  So far, measurements derived from three Lyman limit
absorption systems are available.  One indicates a high value with D/H$=1.9
(\pm 0.5)\times 10^{-4}$ (Songaila et al. 1994; Ruger \& Hogan 1996;
see however Tytler, Burles \& Kirkman 1996), 
and two yields low values with D/H$=2.3 (\pm 0.3\pm 0.3)
\times 10^{-5}$ and $2.5 (\pm 0.5\pm 0.4)\times 10^{-5}$, respectively
(Burles \& Tytler 1996; Tytler, Fan \& Burles 1996). The implications of these
measurements for primordial nucleosynthesis and the value of $\Omega_b$ have
been discussed extensively (Cardall \& Fuller 1996;
Hata {\it et al.} 1997; Jedamzik \& Fuller 1997; Copi, Olive, \& Schramm 1996).
One may be left with an uneasy feeling regarding how representative the
three systems are of Lyman limit systems in general. By necessity only systems
with very narrow lines can show the sought after isotope shift, and only
systems with simple line profiles can minimize the chance of blending. 
Clearly these systems comprise a small fraction of the Lyman limit
systems along the lines of sight to bright QSO's.

The great bulk of deuterium in the universe is generated by freeze-out from
nuclear statistical equilibrium at high entropy in big bang nucleosynthesis
(Wagoner, Fowler, \& Hoyle 1967).  It has been argued effectively that the
bulk of the deuterium cannot be synthesized through cosmic ray spallation or
photo-disintegration of nuclei (e.g., Epstein, Lattimer \& Schramm 1976; Sigl
{\it et al.} 1995).

There are two characteristics of the deuteron which make
its synthesis problematic: (1) it is the most fragile nucleus, with a binding
energy of only $2.2\,{\rm MeV}$; (2) it requires free
neutrons. It is not easy to obtain free neutrons in astrophysical environments,
and to complicate matters, neutrons are unstable. They can only be
\lq\lq spalled\rq\rq\ from bound nuclei, or produced directly via
the weak interaction. The former production channel is difficult because
neutrons are bound in nuclei by $\sim 8\,{\rm MeV}$; while the latter channel
is problematic because weak interactions are inherently slow. Having somehow
produced a free neutron, significant synthesis of deuterons then requires that
the n(p,$\gamma$)D rate be fast compared to the free neutron decay rate and
the local material/thermodynamic expansion timescale,
while the primary deuteron destruction process, D(p,$\gamma$)$^3$He, is slow
compared to the local material/thermodynamic expansion timescale.
Supermassive stars come close to satisfying these requirements
as they are low baryon density, relatively
cool, high entropy configurations, in which the n(p,$\gamma$)D rate
is relatively fast and the D(p,$\gamma$)$^3$He rate is relatively slow.
The free neutrons, in this case, are produced by intense
neutrino fluxes released during the collapse of supermassive stars
via the weak reaction $\bar\nu_e +p\rightarrow n+e^+$.

The crux issue in this scenario for deuterium production hinges on whether {\it
any} material in these stars can be ejected from regions where all the
requirements to yield significant D$/$H enrichment are satisfied. Indeed,
Woosley (1977) suggested that something like this process could account for
all of the deuterium in the universe. Though we will argue that this is not
possible given our current understanding of how supermassive stars evolve
(FWW), it may still lead to significant local deuterium abundance
enrichments, to as high as ${\rm D}/{\rm H} \approx10^{-4}$.  We also show
that the possibility that supermassive star ejecta form a \lq\lq Lyman limit
system\rq\rq\ is consistent with the characteristics of the currently observed
Lyman $\alpha$ absorption systems from which D/H values have been measured.
We then discuss how to eliminate this possibility by further observations.

\section{Deuterium Enrichment in the Envelopes of Supermassive Stars}

In this section we will describe how D$/$H can be enhanced in some regions of a
supermassive star via this three step scenario:
\begin{itemize}
\item The star collapses and, near the point where the homologously collapsing
core becomes a black hole, a fraction of the gravitational binding energy is
radiated as thermally-produced $\nu\bar\nu$ pairs.
\item The process $\bar\nu_e +p\rightarrow n+e^+$ creates free neutrons further
out in regions of the star where material ejection is possible.
\item These neutrons undergo n(p,$\gamma$)D on a timescale short compared to
the material expansion timescale and the neutron decay lifetime; yet the
lifetime of the deuteron against destruction via D(p,$\gamma$)$^3${He} is long
compared to the disassembly time (or inverse expansion rate) of this region of
the star.
\end{itemize}

Leaving aside the thorny and unresolved issue of how a supermassive star could
be formed, it is instructive to review the simple structure and evolution of
nonrotating, primordial-composition supermassive stars (FWW). These stars are
radiation dominated, with essentially all the pressure support coming from
photons, while nearly all the mass-energy in the star is contributed by baryon
rest mass. These stars will have an adiabatic index $\Gamma_1 \approx 4/3$,
will be nearly constant entropy, fully convective configurations, and as a
result will be very accurately described as an index $n=3$ Newtonian polytrope.
Such a configuration will have close to zero total (internal plus Newtonian
gravitational) binding energy and will be \lq\lq trembling on the verge of
instability\rq\rq\ as outlined by Fowler (1964) and Iben (1963).  Additionally,
this configuration will be radiating at the photon Eddington limit.  As a
result, the star will resemble an over-sized Wolf-Rayet star and will be
accompanied by continual and significant mass loss during its lifetime. This
lifetime will be quite short, however, as the star will quasi-statically
contract as it radiates away entropy. This will continue until the central
density reaches a critical value, beyond which general relativistic effects
ensure that the star becomes dynamically unstable. This instability point is
roughly coincident with the onset of hydrogen burning for stars with masses
$\sim {10}^5\,{\rm M_\odot}$. The subsequent fate of the star depends on a
number of factors, principally composition, rotation and magnetic pressure.
Explosion or collapse to a black hole depends on how rapidly thermal energy can
be added to the system through the competition between hot hydrogen burning on
the one hand, and thermal neutrino losses and the build-up of infall kinetic
energy on the other. In general, collapse to a black hole is probably
inevitable for a zero metallicity, nonrotating initial configuration (FWW).

However, there is a generic feature of this collapse which serves to isolate an
outer region of the star from a \lq\lq homologous core\rq\rq\ which will plunge
through the event horizon as a unit. Initially the entire star will collapse
homologously, but as pressure is relatively reduced in the center by thermal
neutrino emission, only an inner part of the star can continue to collapse
homologously (Goldreich \& Weber 1980; FWW). The boundary of this homologous
core (as well as the mass enclosed) depends on the entropy per baryon, which in
turn depends on the history of nuclear burning-driven entropy increase and
neutrino loss-driven entropy decrease (FWW; Fuller \& Shi 1997).
Material outside of this core could be subject to ejection resulting
from neutrino heating (Fuller \& Shi 1997), and/or hot $rp$-process nuclear
burning-driven convection combined with rotation and MHD effects  ({\it cf.},
Ozernoi \& Usov 1971; Fowler 1966). Our work here is meant to suggest
that detailed multi-dimensional numerical modeling of these scenarios merits  
serious attention.

Prodigious thermal neutrino losses will accompany the collapse of the
homologous core (FWW). The peak neutrino-pair production rate will come near
the point where the homologous core of mass ${\rm M}^{\rm HC}$ approaches its
Schwarzschild radius, $r_s \approx 3\times{10}^5\,{\rm cm}
\left({\rm M}^{\rm HC}/{\rm M_\odot}\right)$, where the average density is
$\bar\rho_s \approx 1.8\times{10}^{16}\,{\rm g}\,{\rm cm}^{-3}
{\left({\rm M}^{\rm HC}/{\rm M_\odot}\right)}^{-2}$.
Crudely, the Newtonian gravitational binding energy
at this point is $E_G^{\rm MAX}\approx 0.5{\rm M}^{\rm HC}c^2\approx 
{10}^{54}\,{\rm ergs}\left({{\rm M}^{\rm HC}/{\rm M_\odot}}\right)$, while
the free fall timescale is $t_s \approx 3\times{10}^{-5}\,{\rm s}
\left({{\rm M}^{\rm HC}/{\rm M_\odot}}\right)$.

Not all of the gravitational binding energy will be carried away by neutrinos.
A fraction of the binding energy could reside in infall kinetic energy and
thermal energy, or be dissipated in magnetic modes, or be
\lq\lq stored\rq\rq\ in rotational kinetic energy.  In addition, neutrinos
radiated from the core will be gravitationally redshifted. We designate the
fraction of the binding energy radiated by neutrinos by $f_{\rm E}$.
This quantity $f_{\rm E}$ will be a significant fraction of unity if
neutrinos are not trapped in the homologous core as a result of their mean
free path becoming short compared to the core radius. (Fuller \& Shi 1997 find 
that such trapping is unlikely except very near the black hole formation point
for stars with initial masses above ${\rm M} \ga 10^5\,{\rm M}_\odot$.)

During the collapse the temperature will be $T_9\equiv T/10^9{\rm K}\approx
0.67 {\left( 5.5/g\right)}^{1/3} S_{100}^{1/3}\rho_3^{1/3}$, where $S_{100}$
is entropy per baryon in units of 10$^2$ Boltzmann's constant,
$\rho_3$ is the density in $10^3 {\rm g}\,{\rm cm}^{-3}$
and $g$ is the statistical weight in relativistic particles
($g\approx 2$ and $g \approx 11/2$ for $T_9 \la 1$ and $T_9 \ga 1$,
respectively). $S_{100}$ is typically between 1 to 10.
Near the endpoint of collapse, the core temperature will be of order
$T_9^{\rm HC} \sim 11.7{\left({\rm M}_5^{\rm HC}\right)}^{-1/2}$
where ${{\rm M}_5^{\rm HC}}$ is the homologous core mass in units of
${10}^5\,{\rm M}_\odot$ at the blackhole formation point.
Allowing for entropy loss by thermal neutrino emission
and including a fair amount of angular momentum/centrifugal support
for $M_5\approx 10$ suggests that $M_5^{\rm HC}$ could be about an order
of magnitude smaller than the initial stellar mass (Fuller \& Shi 1997). This
implies that typical temperatures will be $T_9 \sim 10$ to 20 at the
collapse endpoint for initial stellar configurations with
$1\la {\rm M}_5\la 10$.  For $T_9 \sim 10$ to 20 the average energy of
neutrinos produced as pairs in $e^\pm$ annihilation will be
({\it cf.} Schinder {\it et al.} 1987; Itoh et al. 1989)
$\left\langle E_{\nu} \right\rangle \sim 6$ to 12 MeV.  The neutrino
energy will suffer an undetermined gravitational redshift of order 1, but for
simplicity we absorb the redshift into the factor $f_{\rm E}$.

The probability that the process $\bar\nu_e + p\rightarrow n +e^+$  converts
protons at radius $r$ into neutrons is:
\begin{equation}
P\approx {{E_G^{\rm max}f_{\rm E}}\over{\left\langle E_{\nu} \right\rangle}}
{\left({1\over{4\pi r^2}}\right)}{\left({{n_{\bar\nu_e}}\over
{\sum_{i}{n_{\nu_i}}}}\right)} \bigl\langle\sigma(E_\nu)\bigr\rangle.
\end{equation}
The third term in this equation is the fraction of neutrinos which are
$\bar\nu_e$, and is typically about $1/3$. The cross section for
$\bar\nu_e$ absorption on protons is
$\sigma\left( E_\nu\right) \approx {10}^{-43}\,{\rm cm}^2(E_\nu/{\rm MeV})^2$
when $E_\nu\gg 1$ MeV. Therefore,
\begin{equation}
P \approx 1.5\times {10}^{-5}\,f_{\rm E}\,{\rm M^{HC}_5}
\left({{\left\langle E_{\nu}
\right\rangle}\over{10\,{\rm MeV}}}\right)
{\left({{3n_{\bar\nu_e}}\over{\sum_{i}{n_{\nu_i}}}}\right)} 
{\left( {{{10}^{13}\,{\rm cm}}\over{r}} \right)}^{2}.
\end{equation}
With this expression we can estimate that $P \approx 1.5\times{10}^{-5}$ when
$f_{\rm E}\approx 0.1$ and $r=3\times{10}^{12}\, {\rm cm}$ for a homologous
core mass of ${\rm M}_5^{\rm HC}=1$. This is an interesting figure of merit,
as it is roughly the lowest reasonable value for the primordial deuterium
abundance - any production above this value would correspond to D/H
enhancement. A $P$ of 10$^{-4}$ is not out of reach if $f_{\rm E}$ can
be as high as 60$\%$. Note that a baryon at $r=3\times{10}^{12}\, {\rm cm}$
for the indicated value of ${\rm M}_5^{\rm HC}$ would have a
gravitational binding energy of order $0.5\% $ of its rest mass. The key
question is whether there are any circumstances under which material with this
binding energy could acquire enough thermal energy that ejection would be
possible (Fuller \& Shi 1997).

Here $P$ will also be the amount of deuterium enrichment relative to hydrogen
in the limit where the reaction n(p,$\gamma$)D is fast compared to the material
ejection rate and the free neutron decay rate and where the deuteron
destruction process D(p,$\gamma$)$^3${He} is slow compared to the material
ejection rate.  In fact such conditions can be attained for some outer regions
of the supermassive star as we illustrate in Figure 1.  In this figure the
material ejection timescale is taken to be the free-fall timescale, as they
are typically of the same order.
The stippled region corresponds to thermodynamic and expansion conditions
where deuterium can be produced and can survive. 

The fate of the initial primordial deuterium in the envelope is unclear.
Supermassive stars are convective during their pre-collapse
contraction phase.  If their convection proceeds faster than the
contraction, and their central temperatures during most of the contraction
phase are high enough to burn deuterium, the primordial deuterium
may be destroyed in the entire star before collapse.  This could occur in less
massive supermassive stars whose central temperatures are higher.

\section{Envelopes of Supermassive Stars as Lyman $\alpha$ Absorbers}

Once the enriched envelope of a supermassive star is ejected, it expands
freely in space and eventually cools down to
$\sim 10^4$ K, in equilibrium with ambient UV radiation.  Its expansion
velocity is at least $\sim 10$ km/sec, the thermal speed at $10^4$ K.
It can easily expand to a size of 1 kpc,
unless it encounters another cosmic structure or is slowed down simply by
loading of the ambient medium.  The ambient baryonic mass swept up by the
envelope is on average $(4\pi/3)R^3\Omega_b\bar\rho
\sim 10^3\,\Omega_b\,h^2\,(1+z)^3\,(R/1{\rm kpc})^3\,{\rm M}_\odot,$
where $\bar\rho$ is the average matter density at a redshift $z$, and
$R$ is the radius of the expanding envelope. At $z\sim 2$ to 3, this mass
will be larger than the mass of the ejected envelope
if $R\gg 1{\rm kpc}$. Therefore,
the expansion of the envelope up to $R\sim 1$ kpc can be essentially free,
and its chemical composition can be essentially unperturbed.

The number of such envelope systems in one quasar line of sight is
\begin{equation}
(\Omega_{\rm sms}\bar\rho/{\rm M})\pi R^2l (t/t_0)
\sim 300\Omega_{\rm sms}h^2(1+z)^3(R/1{\rm kpc})^2{\rm M}_5^{-1},
\end{equation}
where $\Omega_{\rm sms}$ is the density parameter of baryons in supermassive
stars, and ${\rm M}$ their masses.  $l\sim 10^3$ Mpc is the proper distance
interval under investigation in a quasar line of sight, 
$t\sim 10^8$ years is the duration of the free expansion, and
$t_0\sim$ few $\times 10^9$ years is the age of the universe then.
Therefore, even if we assume that 10$^{-5}$ of the mass
of our universe (or 0.01$\%$ to 0.1$\%$ of baryons) is in these
supermassive stars at $z\sim$ 2 to 3, we can
expect to see at least one such envelope absorber in tens of quasars.
In comparison, there are reported detections from 3 Lyman limit
systems from surveys of $\sim 100$ quasars, a statistic that can easily be
accommodated by these absorbers being ejected envelopes of supermassive stars.

The column densities of these ejected ejected envelopes are approximately
\begin{equation}
x{\rm M}_{\rm env}/\pi R^2m_p\sim 10^{19} x(R/1{\rm kpc})^{-2}
({\rm M}_{\rm env}/10^5{\rm M}_\odot)\,{\rm cm}^{-2},
\end{equation}
where $x$ is the ionization fraction of the envelope, ${\rm M}_{\rm env}$ is
its mass, and $m_p$ is the proton mass. With a temperature of $\sim 10^4$K
and the envelope in photoionization equilibrium with the ambient photoionizing 
UV background, $x\sim 10^{-2}$.
Therefore the resultant column density is that of a Lyman limit system,
$\sim 10^{17}$ cm$^{-2}$, with $M_{\rm env}$ being a significant portion of
the parent mass $\ga 10^5 M_\odot$.

Curiously enough, all three Lyman limit systems with reported
detections of D/H have double components.  This is fully
consistent with the geometry of ejected envelopes.  The velocity separations
between the double components range from 11 to 20 km/sec, which can easily be
produced if the envelopes expand at $\sim 10$ km/sec.
The double components that yield the lower D/H are quite
asymmetric, however.  But one cannot rule out the expanding envelope scenario
on this information alone because so far we have only considered the simplest
case of spherical symmetry.  Asymmetries can easily arise due to asymmetric
ejections or structures in the ambient medium. 
The spectral profiles of the measured clouds indicate temperatures of
$\sim 10^4$K with small turbulence motion.
Our freely expanding envelope scenario fits these features, too.

Explanation of low metalicities in these clouds with measured D/H are
not a problem either, because even though the supermassive stars may undergo
hot hydrogen burning and break-out to the $rp$ process they may not eject
significant amounts of nucleosynthesis products from deep in their cores.
These stars, however, could produce huge amounts of $^4$He (FWW).

Selection criteria dictate that systems with high
deuterium abundances are most likely to be picked out.  Therefore,
if Pop III supermassive stars existed in the early stage of
our universe and they ejected material, all or some of
the 3 Lyman limit systems with detected
deuterium abundances may be accounted for by the ejected envelopes of these
stars.  The required population of these supermassive stars need not
be great. One thousandth of baryons in the universe forming supermassive
stars can be enough.  There is no conflict in having a universe where
$\la 0.1\%$ of its baryonic mass is in $\sim 10^5M_\odot$ black holes
Some of these black hole remnants of supermassive stars may even assume
a role in galaxy formation by accreting mass and becoming the massive black
holes observed at the center of many galaxies (van der Marel et al. 1997).

\section{Discussion}
Thus far we have found that:
\begin{itemize}
\item Anti-electron neutrinos from collapse of supermassive stars can produce
enough free neutrons in outer layers of supermassive stars to boost the
deuterium abundance there to $\sim 10^{-5}$ to 10$^{-4}$.
\item If the deuterium-enriched outer layers get ejected, they can reach a
large enough size that they could
mimic low metalicity Lyman limit absorbers at high redshifts. The chance
of seeing such a Lyman limit absorber in a quasar line of sight is about 1 in
several tens if 0.01$\%$ to $0.1\%$ of baryons are in supermassive stars at
a redshift of 2 to 3, comparable to the current detection probability.
\end{itemize}

Here we especially stress that the ejecta cannot enhance the
global deuterium abundance in any significant manner, because (1) only a small
fraction of baryons may be in the form of supermassive stars (otherwise
most of the baryons now would be in black holes), and (2) only a small
fraction of supermassive star material may be ejected due to the depth of
the gravitational potential. However, we also note that since the
chance of seeing such a local enrichment of deuterium in quasar lines of sight
may not be low, we have to be extremely cautious before we can take a D$/$H
measurement in Lyman limit absorbers as primordial. So how can we assure
ourselves that measurements of D$/$H in Lyman $\alpha$ systems reflect the
primordial value?

First, we do not expect most Lyman limit systems to be envelopes of
supermassive stars. On the other hand, the enrichment process is
sensitive to the conditions in the parent stars, and should not yield
a uniform deuterium abundance.  Consequently, if further observations yield
significantly more detections with a concordant deuterium abundance,
it is highly unlikely that we are measuring enriched values.


Second, if the size of Lyman $\alpha$ absorbers with measured D/H is
established by double quasar lines of sight to be $\gg 1$ kpc,
they cannot be ejecta of supermassive stars.
Their deuterium abundances are therefore not enriched.

Third, the epoch of supermassive stars may not last long.  Thus, observations
by the Hubble Space Telescope of low metalicity absorbers at $z<1$ might
avoid the enrichment problem.  However, one should then also worry about
the destruction problem, that is, whether these absorbers
have been processed through stars and their deuterium has been  
destroyed (Jedamzik \& Fuller 1997).

Ultimately, evidence for supermassive star processing through enhanced D/H
values would suggest that there is some way in which neutrino-exposed ejecta
can escape from fairly deep in the gravitational potential well of these stars,
and the D/H values derived from Lyman limit absorption systems could probe
yet another aspect of structure formation in the early universe.
Along with an enhanced helium abundance, an enhanced deuterium abundance
in Lyman $\alpha$ absorption systems may signal the existence of supermassive
stars at high redshifts.

We thank Scott Burles, Christian Cardall, Karsten Jedamzik,
David Schramm and David Tytler for helpful discussions. The work is supported
by NASA grant NAG5-3062 and NSF grant PHY95-03384 at UCSD.

\newpage
\begin{center}
{\Large Figure Caption}
\end{center}
Figure 1. The plot shows the timescale of n(p,$\gamma$)D and
D(p,$\gamma)^3{\rm He}$, and the dynamic timescale against the temperature
(in units of 10$^9$K) in the star, assuming the density in the star
$\rho\propto T_9^3$. The heavier (lighter) lines are for
${\rm M}_5 \approx 1 (10)$.  The stippled region
corresponds to thermodynamic and expansion conditions where deuterium can be
produced and can survive, for ${\rm M}_5 \approx 1$. The region does not shift
much for ${\rm M}_5 \approx 10$.
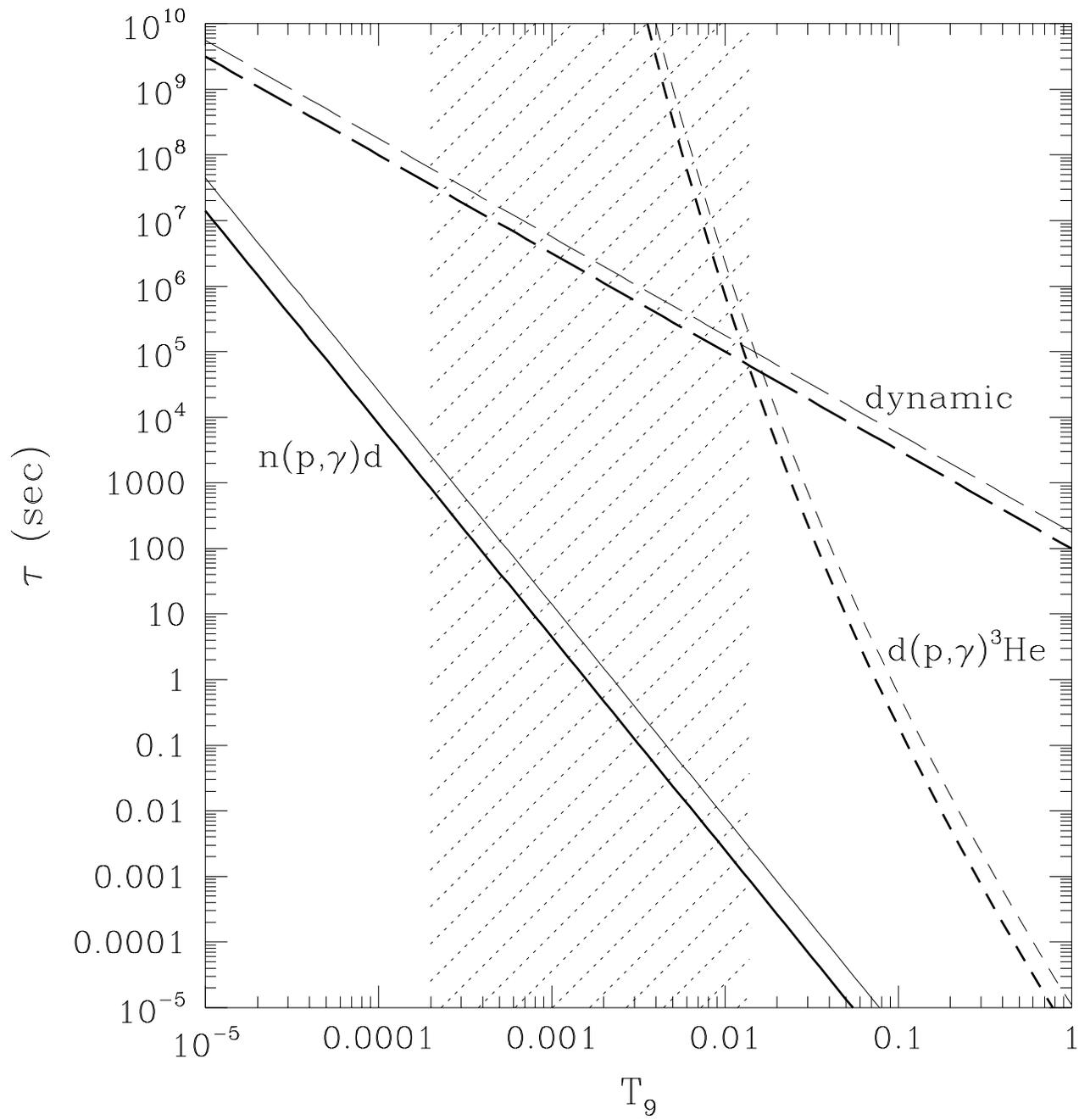
\end{document}